\documentclass[twocolumn,aps,prc,showpacs,floatfix]{revtex4}
\usepackage{color}
\usepackage{graphicx}
\usepackage{dcolumn}
\usepackage{bm}

\begin{document}

\title{Probing the momentum-dependence of the symmetry potential by the free n/p ratio of preequilibrium emission}

\author{He-Lei Liu$^{1,2}$}
\author{Gao-Chan Yong$^{2}$}\email{yonggaochan@impcas.ac.cn}
\author {De-Hua Wen$^{1}$}
\affiliation{%
$^1${School of Science, South China University of Technology,
Guangzhou 510641, P.R. China}\\ $^2${Institute of Modern Physics,
Chinese Academy of Sciences, Lanzhou 730000, China}
}%

\date{\today}

\begin{abstract}
Based on an isospin and momentum-dependent transport model, we
studied the effect of momentum-dependent symmetry potential on the
free neutron to proton ratio of preequilibrium nucleon emission.
It is found that for  $^{132}Sn+^{124}Sn$ reaction at 400
MeV/nucleon incident beam energy, free n/p ratio of preequilibrium
nucleon emission mainly probe the momentum-dependence of symmetry
potential at nucleon momentum around 400 $\sim$ 600 MeV/c. Whereas
for 200 MeV/nucleon incident beam energy, this observable mainly
probe the momentum-dependence of symmetry potential at nucleon
momentum around 200 $\sim$ 400 MeV/c. To probe the symmetry
energy/potential using free n/p ratio, not all the details of the
momentum-dependence of the symmetry potential are important, the
values of symmetry potential at only certain momentum range are
crucial for an observable. It is important to input reasonable
density- and momentum-dependence of the symmetry potential
according to the magnitude of incident beam energy of heavy-ion
collisions. The present experimental data on the symmetry
potential are not enough for probing the density-dependent
symmetry energy. More experimental data (such as nucleon and
nuclei's scattering experiments at different nucleonic momenta and
densities) on the symmetry potential are therefore needed to pin
down the density-dependent symmetry energy.

\end{abstract}


\maketitle

Heavy-ion collisions induced by neutron-rich nuclei have been used
as the unique means to probe the density dependence of nuclear
symmetry energy
\cite{app01,app02,app03,b2005,betty1,betty2,betty3}, the latter is
important to understand the nuclear structures \cite{BAB,CJH} as
well as the properties of neutron stars \cite{JML,KS}. Nowadays,
the density dependence of nuclear symmetry energy/potential is
still not well known especially at supra-saturation densities
\cite{BLC} and high momentum \cite{RBLB}. Few progress has been
made until recently \cite{ZGX}. Also reducing the uncertainties on
the constraints of nuclear symmetry energy is of critical
importance and remains a big challenge.

The momentum-dependence of the symmetry potential is important to
determine the symmetry energy \cite{xuc10,lixh13}. However, except
the old experimental data \cite{DMP}, no new related experiments
on the symmetry potential were carried out in recent 30 years.
Nowadays the determination of the density-dependent symmetry
energy is one of the most important question in nuclear community,
therefore it needs urgently to study the momentum-dependence of
the symmetry potential more detailedly.

Very recently, the authors have done a work of decomposition of
the sensitivity of symmetry energy/potential observables in
density region \cite{HLL} and found that the
symmetry-energy-sensitive observables (including charged pion
ratio $\pi^-/\pi^+$ and free neutron-proton ratio n/p) mainly
probe the symmetry energy/potential around $1.25\rho_0$ (for pion
emission) or $1.5\rho_0$ (for pre-equilibrium nucleon emission).
As a further consideration, for the momentum-dependent symmetry
potential, one would like to know which momentum region does free
n/p ratio reflect. To find out the answer, in this report, we
decompose the sensitivity of the frequently used symmetry energy
observable pre-equilibrium free n/p in momentum region. Such study
is surely crucial to pin down the momentum-dependence of symmetry
potential as well as the density-dependent symmetry energy. In the
following, we show how to decompose the sensitivity of free n/p to
the symmetry potential in momentum space.

\begin{figure}
\centering
\includegraphics [width=0.9\linewidth,clip=true]{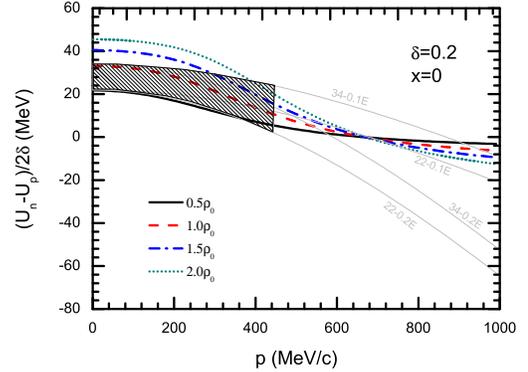}
\caption{\label{potential} (Color online)  Momentum-dependence of
the symmetry potential $(U_n-U_p)/{2\delta}$ at different
densities using the MDI interaction with $x=0$, the shaded region
indicates the experimental constraints.}
\end{figure}
The present study is based on an  isospin and momentum-dependent
transport model IBUU04 \cite{app1,app2}. In the model, the initial
proton and neutron density distributions are given by the
relativistic mean-field theory \cite{ZZR}. We use the experimental
nucleon-nucleon scattering cross section in free space and
consider the Pauli blockings for fermions. The potential adopted
here is an isospin and momentum-dependent single nucleon potential
(MDI) \cite{CBD}, i.e.,
\begin{eqnarray}\label{pot}
U(\rho,\delta,\textbf{p},\tau)=&&A_u(x)\frac{\rho_{\tau'}}{\rho_0}+A_l(x)\frac{\rho_\tau}{\rho_0}\nonumber \\
&&+B(\frac{\rho}{\rho_0})^\sigma(1-x\delta^2)-8x\tau\frac{B}{\sigma+1}\frac{\rho^{\sigma-1}}{\rho_0^\sigma}\delta\rho_{\tau'}\nonumber \\
&&+\frac{2C_{\tau,\tau}}{\rho_0}\int d^3\textbf{p}'\frac{f_\tau(\textbf{r},\textbf{p}')}{1+(\textbf{p}-\textbf{p}')^2/\Lambda^2}\nonumber \\
&&+\frac{2C_{\tau,\tau'}}{\rho_0}\int d^3\textbf{p}'\frac{f_{\tau'}(\textbf{r},\textbf{p}')}{1+(\textbf{p}-\textbf{p}')^2/\Lambda^2},
\end{eqnarray}
where $\rho=\rho_n+\rho_p$ is the baryon density,  $\rho_n$ and
$\rho_p$ denote the neutron $(\tau=1/2)$ and proton $(\tau=-1/2)$
density, respectively. $\delta=(\rho_n-\rho_p)/\rho$ is the
isospin asymmetry of nuclear medium. The variable $x$ is
introduced to reflect different forms of symmetry energy. More
details of the parameters in the above equation can be found in
Refs.~\cite{app1,app2,CBD}. With the above single particle
potential $U(\rho,\delta,\textbf{p},\tau)$, one can give the
symmetry energy $E_{sym}(\rho)$ for a given
parameter $x$. In this study, we choose $x=0$ for example. Fig.~%
\ref{potential} shows the isovector potential
$(U_n-U_p)/{2\delta}$ at different densities using the MDI
interaction with $x=0$. The shaded region can be described by $U=
a-bE_{kin}$ with $a\simeq 22-34 MeV$ and $b\simeq 0.1-0.2$ from
experimental data \cite{DMP,app3,LWC}. Here $E_{kin}$ can be
expressed as a function of nucleonic momentum
$E_{kin}=1000\times[(p^2+0.93893^2)^{1/2}-0.93893]$ MeV (here
nucleonic momentum $p$ is in GeV/c). It is seen that the
experimental data of symmetry potential reach about 450 MeV/c in
momentum space. To better constrain the density-dependent symmetry
energy, it is necessary to extend present experimental data in
large density and momentum region.

\begin{figure}
\centering
\includegraphics [width=0.9\linewidth,clip=true]{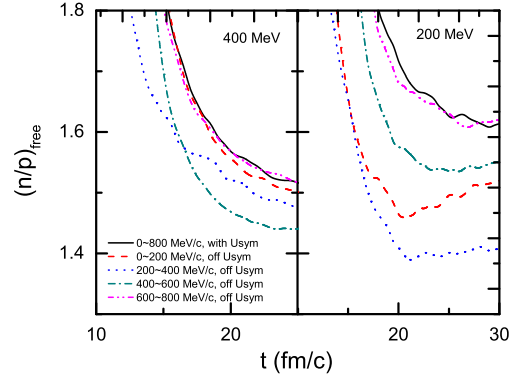}
\caption{\label{snp} (Color online) Evolutions of pre-equilibrium
free n/p ratio $R_{i}$ (see text for details) by turning off the
symmetry potential in momentum regions 0 $<$ p $<$ 200 MeV/c, 200
$<$ p $<$ 400 MeV/c, 400 $<$ p $<$ 600 MeV/c, 600 $<$ p $<$ 800
MeV/c, respectively and comparison with the standard calculation
$R_{0}$.}
\end{figure}
Sensitivity of the free n/p ratio of preequilibrium emission to
the momentum-independent \cite{liba97} and momentum-independent
symmetry potential has been studied many years ago \cite{app1}.
However, the study was just to see the effect of the whole
symmetry potential on the free n/p ratio of preequilibrium
emission. In order to know in which momentum region does the
frequently used symmetry energy/potential sensitive observable
free n/p ratio shows maximal sensitivity to the momentum-dependent
symmetry potential, in the whole momentum region we use
Eq.~(\ref{pot}) with $x = 0$ as the standard calculation, which
gives a value of free n/p ratio $R_{0}$. To see the sensitivity of
this observable to the symmetry potential in momentum space, e.g.,
0 $<$ p $<$ 200 MeV/c, we turn off the symmetry potential in this
momentum space but keep the symmetry potential ($x = 0$) in
residual momentum space. We thus get a new a value of free n/p
ratio $R_{1}$. Similarly we can get other values of free n/p ratio
$R_{2}, R_{3}, ..., R_{i}$ corresponding to other momentum regions
(evolutions of $R_{i}$ are shown in Fig.~\ref{snp}). Through
comparing these new computing results at each step with the
standard calculation, one can obtain the relative sensitivity (=
$\frac{|R_{0}-R_{i}|}{R_{0}} $) of a symmetry energy/potential
observable in certain momentum region.

\begin{figure}
\centering
\includegraphics [width=0.9\linewidth,clip=true]{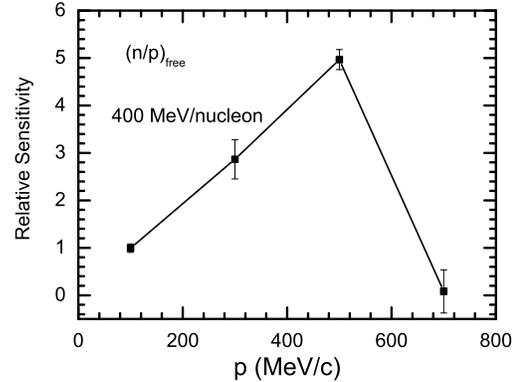}
\caption{\label{np} Relative sensitivity of pre-equilibrium free
n/p ratio to the symmetry potential in different momentum regions
at the beam energy of 400 MeV/nucleon in $^{132}Sn+^{124}Sn$
collision with an impact parameter b=1 fm.}
\end{figure}
Shown in Fig.~\ref{np} is the relative sensitivity of
pre-equilibrium free n/p ratio to the symmetry potential in
different momentum regions at the beam energy of 400 MeV/nucleon
in $^{132}Sn+^{124}Sn$ collision with an impact parameter b=1 fm.
One can clearly see that the maximal sensitivity of n/p ratio to
the symmetry potential is in 400$\sim$ 600 MeV/c momentum space.
Above 600 MeV/c or below 300 MeV/c, free n/p ratio to the symmetry
potential in momentum space is no longer sensitive. Therefore, to
study the effect of nuclear symmetry energy using heavy-ion
collisions at incident beam energies around 400 MeV/nucleon, the
values of symmetry potential in momentum range 400 $\sim$ 600
MeV/c play crucial role in determination the value of the final
observable.

\begin{figure}
\centering
\includegraphics [width=0.9\linewidth,clip=true]{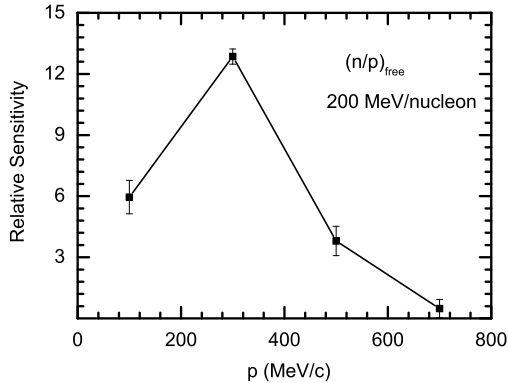}
\caption{\label{mom} Same as Fig.~\ref{np}, but at the beam energy
of 200 MeV/nucleon.}
\end{figure}
Since related experiments with incident beam energies around 100
$\sim$ 350 MeV/nucleon are planning/doing at facilities that offer
fast radioactive beams such as the National Superconducting
Cyclotron Laboratory (NSCL) and the Facility forRare Isotope Beams
(FRIB) in the USA, the Radioactive Isotope Beam Facility (RIBF) in
Japan, it is necessary to make a similar study at incident beam
energy below 400 MeV/nucleon. Shown in Fig.~\ref{mom} is the
relative sensitivity of pre-equilibrium free n/p ratio to the
symmetry potential in different momentum regions at beam energy of
200 MeV/nucleon. However, one can see that the maximal sensitivity
of free n/p ratio to the symmetry potential is in the momentum
space 200 $\sim$ 400 MeV/c. Below the nucleon momentum 200 MeV/c,
the sensitivity is about one half of the maximal sensitivity. When
nucleon momentum is larger than 500 MeV/c, the relative
sensitivity rapidly decrease with nucleonic momentum. More
interestingly, one can see that the sensitivity of free n/p ratio
is about one time larger than that for 400 MeV/nucleon beam
energy. Therefore, to study the effect of nuclear symmetry energy
using heavy-ion collisions at incident beam energies around 200
MeV/nucleon, the values of symmetry potential in momentum range
200 $\sim$ 400 MeV/c play crucial role in determination the value
of the final observable. At least, the momentum-dependent symmetry
potential around saturation density used in the transport model
must fit the experimental data well (as shaded area shown in Fig.~%
\ref{potential}).

\begin{figure}
\centering
\includegraphics [width=0.9\linewidth,clip=true]{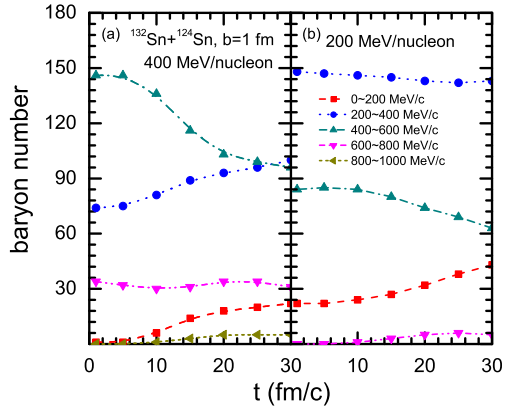}
\caption{\label{NT} (Color online) Location of baryon number in
momentum space as a function of time in $^{132}Sn+^{124}Sn$
reactions at incident beam energies of 400 (a) and 200 MeV/nucleon
(b), respectively.}
\end{figure}
To better understand why the free n/p ratio shows different
sensitivity to the symmetry potential in momentum space, shown in
Fig.~\ref{NT}, we provide the evolution of baryon number at
different momentum regions. It is seen that the distribution of
baryons at different momentum regions changes with reaction time.
For 400 MeV/nucleon incident beam energy, in the whole reaction
process, baryon number in the region of 400 $\sim$ 600 MeV/c is
evidently larger than that located in other regions. While the
maximal baryon number locates in the region of 200 $\sim$ 400
MeV/c at the incident beam energy of 200 MeV/nucleon. Maximal
baryon number location in certain momentum space should be also
maximally affected by the symmetry potential at that momentum
space, thus we see different sensitivity of free n/p ratio to
symmetry potential in momentum space.

It is known to all that the symmetry energy has kinetic and
potential parts \cite{hen14}. However, in transport model
simulations of heavy-ion collisions, the symmetry potential
corresponding to a given potential symmetry energy is a direct
input but the kinetic symmetry energy does not directly enter
transport model simulations \cite{liba14}. Because observables are
sensitive to the symmetry potential mainly at certain density and
momentum ranges, to probe the symmetry energy using heavy-ion
collisions, it is important to input reasonable density- and
momentum-dependence of the symmetry potential \cite{app2,RBLB}
according to the magnitude of incident beam energy of heavy-ion
collisions.

In summary, based on the IBUU04 transport model, we decomposed the
sensitivity of pre-equilibrium free n/p ratio to the
momentum-dependent symmetry potential in momentum region. It is
found that the free n/p ratio mainly probe the symmetry potential
in the momentum region 400 $\sim$ 600 MeV/c for the incident beam
energy of 400 MeV/nucleon and 200 $\sim$ 400 MeV/c for 200
MeV/nucleon beam energy. This study can help us to constrain the
momentum-dependence of the symmetry potential more detailedly.
Since observables are sensitive to the symmetry potential mainly
at certain density and momentum ranges, to probe the symmetry
energy using heavy-ion collisions, it is important to input
reasonable density- and momentum-dependence of the symmetry
potential according to the magnitude of incident beam energy of
heavy-ion collisions. Our study also shows that, the present
experimental data on symmetry potential are not enough for probing
the density-dependent symmetry energy. More experimental data
(such as nucleon and nuclei's scattering experiments at different
nucleonic momenta and densities) on the symmetry potential are
needed to pin down the density-dependent symmetry energy.


This work is supported in part by the National Natural Science
Foundation of China under Grant Nos. 11375239, 11435014, 11275073
and the Fundamental Research Funds for the Central University of
China under Grant No. 2013ZG0036.

\end{document}